\documentclass[journal]{IEEEtran}
\IEEEoverridecommandlockouts

\usepackage{cite}
\usepackage{amsmath,amssymb,amsfonts}
\usepackage{algorithm}
\usepackage{algpseudocode}
\usepackage{graphicx}
\usepackage{caption}
\usepackage{subcaption}
\usepackage[compact]{titlesec}  
\titlespacing{\section}{2pt}{2pt}{2pt}
\usepackage{textcomp}
\usepackage{xcolor}
\usepackage{comment}
\usepackage{hyperref}
\def\BibTeX{{\rm B\kern-.05em{\sc i\kern-.025em b}\kern-.08em
    T\kern-.1667em\lower.7ex\hbox{E}\kern-.125emX}}
\usepackage{bm}
\usepackage{enumitem}
\usepackage[normalem]{ulem}
\usepackage{cases}
\begin{document}

\title{Wireless Federated Learning over UAV-enabled Integrated Sensing and Communication 
\thanks{}
}



\author{
        \IEEEauthorblockN{
        Shaba Shaon\IEEEauthorrefmark{2}, Tien Nguyen\IEEEauthorrefmark{4}, Lina Mohjazi\IEEEauthorrefmark{3}, Aryan Kaushik\IEEEauthorrefmark{5}, Dinh C. Nguyen\IEEEauthorrefmark{2} 
	}

	\IEEEauthorblockA{
	\IEEEauthorrefmark{2}Department of Electrical and Computer Engineering, University of Alabama in Huntsville, USA \\
    \IEEEauthorrefmark{3}James Watt School of Engineering, University of Glasgow, UK \\
    \IEEEauthorrefmark{5}Department of Computing \& Mathematics, Manchester Metropolitan University, UK \\
    \IEEEauthorrefmark{4}Department of Electrical and Electronics Engineering, Lac Hong University, Vietnam
    \\
    Emails: ss0670@uah.edu, chitien2802@gmail.com,  lina.mohjazi@glasgow.ac.uk, a.kaushik@ieee.org, dinh.nguyen@uah.edu
	}\vspace{-0pt}}
	\markboth{}%
	{}
\maketitle

\begin{abstract}
This paper studies a new latency optimization problem in unmanned aerial vehicles (UAVs)-enabled federated learning (FL) with integrated sensing and communication. In this setup, distributed UAVs participate in model training using sensed data and collaborate with a base station (BS) serving as FL aggregator to build a global model. The objective is to minimize the FL system latency over UAV networks by jointly optimizing UAVs' trajectory and resource allocation of both UAVs and the BS.  The formulated optimization problem is troublesome to solve due to its non-convexity. Hence, we develop a simple yet efficient iterative algorithm to find a high-quality approximate solution, by leveraging block coordinate descent and successive convex approximation techniques. Simulation results demonstrate the effectiveness of our proposed joint optimization strategy under practical parameter settings, saving the system latency up to 68.54\% compared to benchmark schemes.

\end{abstract}
\section{Introduction}
Unmanned aerial vehicles (UAVs), commonly referred to as drones, have been revolutionizing next-generation wireless networks with their versatile capabilities including line-of-sight (LoS) connections, 3D mobility, and flexibility. UAVs can function as airborne base stations (BSs) for delivering communication, computation, and caching services to overcome traditional infrastructure limitations, while can also serve as mobile users for tasks like remote sensing, delivery services, target tracking, and virtual reality support. More recently, UAVs have been integrated with machine learning (ML) to deliver intelligent services such as classification of aerial images captured by UAV's cameras. To ensure data privacy during ML model training in UAV networks, federated learning (FL) has been recently employed where UAVs can train an ML model and only share the trained model updates to a cloud server without data exchange. 

Previous studies have mainly focused on implementing conventional FL architectures with UAVs, where UAVs typically serve as FL clients with pre-stored datasets or as aerial base stations responsible for model aggregation. More recent research, such as \cite{6,9}, has integrated UAV deployment into the FL system, enabling UAVs to function as relays for efficient data collection from other devices or as aggregators with adaptive positioning to improve FL performance. Furthermore, much of the prior work in this domain has primarily concentrated on communication and/or computation aspects, assuming that the data for training is readily available without accounting for the data sensing process. This assumption overlooks a critical factor, as the sensing process competes with communication and computation for limited resources, which can significantly affect overall learning performance. In FL, sensing, computation, and communication are highly coupled and must be seamlessly integrated to fully unlock its potential. This has given rise to the increasingly prominent research field of integrated sensing and communication (ISAC) \cite{kaushik2024toward}. 

Several studies have explored FL-UAV, FL-ISAC networks. In \cite{5}, the authors investigated a UAV-FL system for image classification in area exploration scenarios, while \cite{11} proposed UAV-empowered wireless power transfer to enable sustainable FL-based wireless networks. The works in \cite{12}, \cite{3}, \cite{10} studied FL-ISAC frameworks where they focus on jointly optimizing sensing, communication, and computation resource allocation. Despite these research efforts, \textit{the problem of latency minimization in UAV-enabled FL system with ISAC remains under-explored}. Given UAV's limited computational power and battery capacity, optimizing the round-trip ML model training latency, such as ML model training and ML model communication, is crucial to achieve efficient and timely FL while preserving UAVs' resources.

Hence, \textit{this paper studies a new latency minimization problem for UAV-enabled FL with ISAC}. The contributions of this paper are two-fold: (1) We formulate a new latency minimization problem for UAV-enabled FL with ISAC, by jointly optimizing resource allocation and trajectory of UAVs along with resource allocation of the BS (Section II); (2) This latency minimization problem is computationally challenging, thus we propose a new iterative optimization approach to obtain the optimal solutions by applying block coordinate descent (BCD) and successive convex approximation (SCA) techniques (Section III). Numerical simulations show the superior performance of our joint optimization method compared to baseline approaches (Section IV). 

\section{System Model and Problem Formulation}

\subsection{System Model}
We consider a UAV-ISAC scenario where distributed UAVs collaboratively train a shared ML model  with a base station (BS) that functions as a model aggregator. The set of UAVs is denoted by $\mathcal{N} = \{1,2,\dots,N\}$, where each UAV $n$ is equipped with a single-antenna transceiver that can alternate between sensing and communication modes as required. This switching occurs in a time-division manner using a shared radio-frequency front-end circuit \cite{zhang2022accelerating}. 


Each UAV $n$ is assumed to sense data $\mathcal{D}_{n}$ with size $D_{n} \triangleq |\mathcal{D}_{n}|$. Moreover, we denote the union of datasets collected across all the UAVs, called global dataset, as $\mathcal{D} = \cup_{n \in \mathcal{N}} \mathcal{D}_{n}$ with size $D \triangleq |\mathcal{D}|$. The procedure of the considered framework in each global FL round is summarized as follows:
\begin{enumerate}
    \item Each of participating UAVs senses data from the ground objects within its coverage area. 
    \item Using the initial global model shared by the BS, each UAV trains its local model based on sensed data.
    \item Each UAV transmits its updated local model to the BS.
    \item The BS aggregates the received updated local models to build a new version of the global model.
    \item The BS broadcasts the aggregated global model to the participating UAVs for the next round of training.   
\end{enumerate}


We consider a three-dimensional (3D) Cartesian coordinate system where ground objects and the BS are static, with the BS at coordinates $(0,0,0)$. Each UAV will hover overhead to sense data and train its local model. Then it flies toward the BS to upload the updated local model. To facilitate the analysis, we assume that UAV $n$ flies from its initial position $(x_{n}^{0},y_{n}^{0},H)$ to its final position $(x_{n}^{f},y_{n}^{f},H)$ at a fixed altitude $H > 0$ above the ground during its operation time $T_{\text{flight}}$. This is the communication period for local model uploading and is discretized into $T$ equal time slots, where each slot is given by $\delta_{t} = \frac{T_{\text{flight}}}{T}$. To guarantee the approximate invariability of the UAV’s location within each slot, the slot length is chosen to be sufficiently small. Therefore, the dynamic horizontal coordinate can be defined by $q_{n}[t] \triangleq (x_{n}[t],y_{n}[t])$, $t \in \mathcal{T} \triangleq \{1,2,\dots,T\}$. Let $V_{\max}$ denote the maximum UAV speed. Then we have $\hat{q}_{n}(t_{\text{flight}}) \leq V_{\max}$, $0 \leq t_{\text{flight}} \leq T_{\text{flight}}$, where $\hat{q}_{n}(t_{\text{flight}})$ denotes the first derivative of the UAV's location with respect to $t_{\text{flight}}$. Consequently, we have the constraint on the UAV trajectory as $(x_{n}[t+1] - x_{n}[t])^2 + (y_{n}[t+1] - y_{n}[t])^2 \leq (V_{\max} \delta_{t})^2$, 
where $V_{\max} \delta_{t}$ represents the maximum UAV displacement during each time slot $t$. It is
assumed that the single-antenna ground objects cannot be served by the BS directly due to the blockage of surrounding obstacles. We further assume that all the links between UAVs and BS, as well as those between UAVs and their sensing targets, are line-of-sight (LoS) channels. Thus, the LoS channel gain between UAV $n$ and BS at time slot $t$ follows free space pathloss model, which can be given by $g_{n,\text{BS}}^{(k)}[t] = \frac{\beta_{0}}{d_{n,\text{BS}}^{(k)}[t]^2}$, where $\beta_{0}$ represents channel gain at reference distance $d_{0}=1 m$, and $d_{n,\text{BS}}$ denotes distance from UAV $n$ to BS at time slot $t$.

Each UAV $n$ has a model $\boldsymbol{w}_{n}$ for its dataset $\mathcal{D}_{n}$, which is designed to capture the relationship between the input data and its corresponding label. Each model, when trained on the dataset, yields a predicted label denoted as $\hat{y}_{n} = \boldsymbol{w}_{n}(\mathcal{D}_{n})$, where $\hat{y}_{n}$ represents the predicted label generated from model $\boldsymbol{w}_{n}$ for dataset $\mathcal{D}_{n}$. We denote the local loss function for UAV $n$ as $\mathcal{L}_{n}(\boldsymbol{w}_{n};\mathcal{D}_{n}) = \frac{1}{|\mathcal{D}_{n}|} \sum_{d \in \mathcal{D}_{n}} \ell(\boldsymbol{w}_{n};d),$
where $\ell(\boldsymbol{w}_{n};d)$ is the ML model loss function evaluated on data point $d$. Subsequently, the global loss function can be denoted as $\mathcal{L}(\boldsymbol{w}) \triangleq \sum_{n \in \mathcal{N}} \frac{|\mathcal{D}_{n}|}{|\mathcal{D}|} \mathcal{L}_{n}(\boldsymbol{w}_{n};\mathcal{D}_{n}),$
where $\boldsymbol{w}$ is defined as global model. The local model training at each UAV is carried out through a series of minibatch stochastic gradient descent (SGD) iterations. We assume that each UAV $n$ participates in $S$ SGD iterations and $K$ global communication rounds. Specifically, given a local model at UAV $n$, at SGD iteration $s$, at global round $k$ (i.e., $\boldsymbol{w}_{n}^{k, s}$), the next local model is obtained as $\boldsymbol{w}_{n}^{k,\text{s+1}} = \boldsymbol{w}_{n}^{k,s} - \eta^{k,s} \Tilde{\nabla} \mathcal{L}_{n}(\boldsymbol{w}_{n}^{k,s};\beta_{n}^{k,s})$, 
where $\Tilde{\nabla} \mathcal{L}_{n}(\boldsymbol{w}_{n}^{k,s};\beta_{n}^{k,s}) \triangleq \frac{1}{\beta_{n}^{k,s}} \sum_{d \in \beta_{n}^{k,s}} \nabla \ell(\boldsymbol{w}_{n}^{k,s};d)$.
Here, $\beta_{n}^{k,s}$ denotes a mini-batch of data sampled uniformly at random from $\mathcal{D}_{n}$ and $\eta^{k,s}$ is the SGD learning rate. As UAV $n$ conducts $S$ SGD iterations before transmitting its updated model to the server, we let $\boldsymbol{w}_{n}^{k,S}$ denote the final model at UAV $n$. After local training, BS receives updated models from all the participating UAVs and performs aggregation of the models as $\boldsymbol{w}_{l}^{\text{k+1}} = \frac{\sum_{n=0}^{N} |\mathcal{D}_{n}| \boldsymbol{w}_{n}^{k,S}}{|\mathcal{D}|}, \forall l \in \mathcal{L}$. 
The latency for UAV $n$ in communication round $k$ comprises five parts: data sensing, local model training, local model uploading, global aggregation, and global model downloading. We explicitly formulate each of these latency components as below.

\begin{enumerate}[label=\textit{\pmb{\arabic*}})]
    \item \textit{\pmb{Data sensing time:}} Each ISAC-enabled UAV performs radar sensing within its coverage area to capture radar echo signal reflected from ground targets. This signal is converted into a set of data bits used for local model training. Let us denote the unit sensing time, i.e., the time for generating a sensing sample, as $T_{u}$. If the number of samples generated by UAV $n$ at communication round $k$ is $D_{n}^{(k)}$, its data sensing time at round $k$ is given by \cite{3}
    \begin{align}
        T_{\text{se},n}^{(k)} = T_{u} D_{n}^{(k)}.
    \end{align}
    The corresponding UAV energy consumption is
    \begin{align}
        E_{\text{se},n}^{(k)} = p_{\text{se},n}^{(k)} T_{\text{se},n}^{(k)},
    \end{align}
    where $p_{\text{se}, n}^{(k)}$ is the sensing transmit power of UAV $n$ at round $k$.
    \item \textit{\pmb{Local model training time:}} As the number of local iterations for UAV $n$ at round $k$ is $S$, its local computation time at round $k$ can be expressed as
    \begin{align}
        T_{\text{train},n}^{(k)} = \frac{S C_{n}^{(k)} D_{n}^{(k)}}{f_{n}^{(k)}},
    \end{align}
    where $C_{n}^{(k)}$ is the number of CPU cycles needed to execute computation of a sample in a local update and $f_{n}^{(k)}$ is the CPU computation capability (in cycles/s). The corresponding UAV energy consumption can be given by
    \begin{align}
        E_{\text{train},n}^{(k)} = S \zeta_{n}^{(k)} C_{n}^{(k)} D_{n}^{(k)} \left(f_{n}^{(k)}\right)^2,
    \end{align}
    where $\zeta_{n}^{(k)}$ is effective switched capacitance that depends on the hardware and chip architecture of the UAV \cite{6}.
    \item \textit{\pmb{Local model uploading time:}} For ease of calculation, we assume that the local model size that each UAV needs to upload in every communication round at time slot $t$ is constant, which is denoted by $s_{l}[t]$. At time slot $t$, the local model uploading time of UAV $n$ at round $k$ is \vspace{-10pt}
    \begin{align}
        T_{\text{cm},n}^{(k)} = \frac{s_{l}[t]}{R_{n}^{(k)}[t]},
    \end{align}
    where $R_{n}^{(k)}[t]$ is the uplink transmission rate of UAV $n$ to BS at round $k$ at time slot $t$, which can be given by \cite{2} \vspace{-12pt}
    \begin{align}
        R_{n}^{(k)}[t] &= B_{n} \log_{2}\left(1+\frac{g_{n, \text{BS}}^{(k)}[t] p_{\text{cm},n}^{(k)}[t]}{\sigma^{2}}\right) \notag \\
        &= B_{n} \log_{2}\left(1+\frac{\beta_{0} p_{\text{cm},n}^{(k)}[t]}{\sigma^{2} \left(d_{n,\text{BS}}^{(k)}[t]\right)^2}\right) \notag \\
        &= B_{n} \log_{2}\left(1+\frac{\gamma_{0} p_{\text{cm},n}^{(k)}[t]}{\left(d_{n,\text{BS}}^{(k)}[t]\right)^2}\right),
    \end{align}
    where $\gamma_{0} = \frac{\beta_{0}}{\sigma^{2}}$ is the reference signal-to-noise ratio (SNR). Further, $B_{n}$ is the communication bandwidth allocated for UAV $n$, $p_{\text{cm},n}^{(k)}[t]$ is the communication transmit power of UAV $n$ at round $k$ at time slot $t$, $\sigma^{2}$ is the additive white Gaussian noise (AWGN) power at the BS, and $d_{n,\text{BS}}^{(k)}[t]$ is the LoS distance between UAV $n$ and BS at round $k$ at time slot $t$ which can be given by
    \begin{equation}
    d_{n,\text{BS}}^{(k)}[t] = \sqrt{(x_{n}^{(k)}[t])^{2}+(y_{n}^{(k)}[t])^{2}+H^{2}}.
    \end{equation}
    The UAV energy consumption for T time slots is \vspace{-8pt}
    \begin{align}
        E_{\text{cm},n}^{(k)} = \sum_{t=1}^{T}T_{\text{cm},n}^{(k)} p_{\text{cm},n}^{(k)}[t].
    \end{align}
    \item \textit{\pmb{Global aggregation time:}} The BS aggregates uploaded local models. Accordingly, the global aggregation time at round $k$ can be written as \vspace{-8pt}
    \begin{align}
        T_{\text{cp},\text{BS}}^{(k)} = \frac{C_{\text{BS}}^{(k)} D_{\text{BS}}^{(k)}}{f_{\text{BS}}^{(k)}}, 
    \end{align} 
    \vspace{-8pt}
    
    where $C_{\text{BS}}^{(k)}$ is the number of CPU cycles needed to execute the computation of a sample and $f_{\text{BS}}^{(k)}$ is the CPU computation capability (in cycles/s). 
    \item \textit{\pmb{Global model downloading time:}} The BS broadcasts the updated global models to all the UAVs over the entire frequency band. If $R_{\text{BS}}^{(k)}$ is the downlink transmission rate of BS to UAV $n$ at round $k$, then it can be formulated as
    \begin{align}
        R_{\text{BS}}^{(k)} &= B_{\text{BS}}\log_{2}\left(1+\frac{g_{\text{BS},n}^{(k)} p_{\text{cm},\text{BS}}^{(k)}}{\sigma^{2}}\right) \notag \\
        &= B_{\text{BS}} \log_{2}\left(1+\frac{\beta_{0} p_{\text{cm},\text{BS}}^{(k)}}{\sigma^{2} \left(d_{\text{BS},n}^{(k)}\right)^2}\right) \notag \\
        &= B_{\text{BS}} \log_{2}\left(1+\frac{\gamma_{0} p_{\text{cm},\text{BS}}^{(k)}}{\left(d_{n,\text{BS}}^{k}[t]\right)^2}\right),
    \end{align} 
    where $B_{\text{BS}}$ is the communication bandwidth that the BS uses to broadcast the global model to each UAV, $g_{\text{BS},n}^{(k)}$ is the LoS channel gain between BS and UAV $n$ at round $k$ which is same as $g_{n,\text{BS}}^{(k)}[t]$, $p_{\text{cm},\text{BS}}^{(k)}$ is the communication transmit power of BS at round $k$, $\sigma^{2}$ is the AWGN power at UAV $n$, $\gamma_{0} = \frac{\beta_{0}}{\sigma^{2}}$ is the reference signal-to-noise ratio (SNR), and $d_{\text{BS},n}^{(k)}$ is the LoS distance between BS and UAV $n$ at round $k$ which is same as $d_{n,\text{BS}}^{k}[t]$. If we denote global model size as $s_{g}$, global model downloading time of UAV $n$ at round $k$ can be formulated as
    \begin{align}
        T_{\text{dl},n}^{(k)} = \frac{s_{g}}{R_{\text{BS}}^{(k)}}.
    \end{align}    
\end{enumerate}

In this work, we consider synchronous aggregation at the BS, i.e., the aggregation does not happen until the local models from all the participating UAVs arrive at the BS. Hence, the time consumption of any communication round $k$ is determined by the most time-consuming UAV, i.e., \vspace{-5pt}
\begin{align}
\footnotesize
    T^{(k)} = \underset{n \in \mathcal{N}}{\max}\left\{T_{\text{se},n}^{(k)}+T_{\text{train},n}^{(k)}+T_{\text{cm},n}^{(k)}+T_{\text{cp},\text{BS}}^{(k)}+T_{\text{dl},n}^{(k)}\right\}.
\end{align}
Accordingly, the total FL latency can be written as 
\begin{align}
    T_{\text{total}}^{\text{FL}} = \sum_{k=1}^{K} T^{(k)},
\end{align}
where $K$ is the total number of communication rounds in the FL training process. In terms of energy consumption, the BS generally has much energy resources, while UAVs are energy-constrainted devices. Hence, we focus on energy formulation for UAV, which is expressed as
\begin{align}
    E_{\text{total},n}^{\text{UAV}} = \sum_{k=1}^{K}\left(E_{\text{se},n}^{(k)}+E_{\text{train},n}^{(k)}+E_{\text{cm},n}^{(k)}\right). \label{eqn:18}
\end{align}
From \eqref{eqn:18}, the energy consumption of each UAV comes from three parts: energy for sensing, energy for local computation, and energy for local model uploading. The BS is typically equipped with substantial battery reserves or even connected to the power grid, making it a highly reliable infrastructure element. 

\subsection{Problem Formulation}
This work aims to minimize the latency of the UAV-ISAC FL system. Based on the above analysis, we formulate a system latency optimization problem which aims to joinly optimize UAV trajectory ($x_{n}^{(k)}[t]$, $y_{n}^{(k)}[t]$) and resource allocation of both UAV ($p_{\text{se},n}^{(k)}$, $p_{\text{cm},n}^{(k)}[t]$, $f_{n}^{(k)}$) and the BS ($p_{\text{cm},\text{BS}}^{(k)}$ and $f_{\text{BS}}^{(k)}$).
\\ 
\textit{\pmb{Problem 1:}} \vspace{-13pt}
\begin{subequations} 
\begin{align}
\min_{\begin{aligned}
    & \text{\small $x_{n}^{(k)}[t]$, $y_{n}^{(k)}[t]$,} \\[-0.5ex]
    & \text{\small$p_{\text{se},n}^{(k)}$, $p_{\text{cm},n}^{(k)}[t]$,} \\[-0.5ex]
    & \text{\small $p_{\text{cm},\text{BS}}^{(k)}$, $f_{n}^{(k)}$, $f_{\text{BS}}^{(k)}$} 
\end{aligned}} \quad & T_{\text{total}}^{\text{FL}} \tag{15a} \label{eqn:15a}\\ 
\textrm{s.t.} \quad & 0 \le p_{\text{se},n}^{(k)} \le P_{\text{se},n}^{\max}, \forall{k,n} \tag{15b} \label{eqn:15b}\\
& 0 \le p_{\text{cm},n}^{(k)}[t] \le P_{\text{cm},n}^{\max}, \forall{k,n,t} \tag{15c} \label{eqn:15c}\\
& 0 \le p_{\text{cm},\text{BS}}^{(k)} \le P_{\text{cm},\text{BS}}^{\max}, \forall{k} \tag{15d} \label{eqn:15d}\\
& 0 \le f_{n}^{(k)} \le f_{n}^{\max}, \forall{k,n} \tag{15e} \label{eqn:15e}\\
& 0 \le f_{\text{BS}}^{(k)} \le f_{\text{BS}}^{\text{max}}, \forall{k} \tag{15f} \label{eqn:15f}\\
& E_{\text{total},n}^{\text{UAV}} \le E^{\max}, \forall{n} \tag{15g} \label{eqn:15g}\\
 (x_{n}[t+1] - x_{n}[t])^2 &+ (y_{n}[t+1] - y_{n}[t])^2 \leq (V_{\max} \delta_{t})^{2}. \tag{15h} \label{eqn:15h}
\end{align} 
\end{subequations} \vspace{-15pt}
\\
where $p_{\text{se},n}^{(k)} = \{p_{\text{se},1}^{(k)},p_{\text{se},2}^{(k)},\dots,p_{\text{se},N}^{(k)}\}$, $p_{\text{cm},n}^{(k)}[t] = \{p_{\text{cm},1}^{(k)}[t],p_{\text{cm},2}^{(k)}[t],\dots,p_{\text{cm},N}^{(k)}[t]\}$, and $f_{n}^{(k)}=\{f_{1}^{(k)},f_{2}^{(k)},\dots,f_{n}^{(N)}\}$. In the above problem, $P_{\text{se},n}^{\max}$, $P_{\text{cm},n}^{\max}$,  $f_{n}^{\max}$ are maximum sensing transmit power, maximum communication transmit power, and maximum CPU computation capability of UAV $n$, respectively. $P_{\text{cm},\text{BS}}^{\max}$, $f_{\text{BS}}^{\max}$ are maximum communication transmit power and maximum CPU computation capability of BS. $E^{\max}$ is the maximum energy consumption of a UAV. Here, \eqref{eqn:15b} and \eqref{eqn:15c} represent the feasible ranges of the sensing and the communication transmit power, respectively, due to the power budgets of UAVs. \eqref{eqn:15d} implies the constraint of the BS's transmit power. The CPU computation capability of each UAV and BS are constrained in \eqref{eqn:15e} and \eqref{eqn:15f}, respectively. \eqref{eqn:15g} is on the maximum energy consumption by UAV $n$. \eqref{eqn:15h} restricts the maximum distance UAV $n$ is allowed to travel within time slot $t$.

\section{Proposed Solution}
\pmb{Problem 1} is hard to solve directly due to the non-convex nature of the objective function as well as the constraints. In order to solve \pmb{Problem 1}, based on the BCD technique, we divide the original optimization problem into two sub-problems. Hence, the control variables of problem 1 are divided into two blocks: UAV parameters optimization $(x_{n}^{(k)}[t], y_{n}^{(k)}[t], p_{\text{se},n}^{(k)}, p_{\text{cm},n}^{(k)}[t], f_{n}^{(k)})$ and BS parameters optimization $(p_{\text{cm},\text{BS}}^{(k)}, f_{\text{BS}}^{(k)})$. Then, we acquire the solution to the original problem by solving the two sub-problems iteratively until the objective function is convergent.
\\
\textit{\pmb{Sub-problem 1:}}
\vspace{-15pt}
\begin{subequations} 
\begin{align}
\min_{\begin{aligned}
    & \text{\small $x_{n}^{(k)}[t]$, $y_{n}^{(k)}[t]$,} \\[-0.5ex]
    & \text{\small$p_{\text{se},n}^{(k)}$, $p_{\text{cm},n}^{(k)}[t]$,} \\[-0.5ex]
    & \text{\small $f_{n}^{(k)}$}
\end{aligned}} \quad & \sum_{k=1}^{K}\max_{n \in \mathcal{N}} \left\{T_{u} D_{n}^{(k)}+S C_{n}^{(k)} D_{n}^{(k)}/f_{n}^{(k)}\right. \nonumber \\[-30pt]
& \left.+s_{l}[t]/R_{n}^{(k)}[t]+T_{\text{cp},\text{BS}}^{(k)}+T_{\text{dl},n}^{(k)}\right\} \tag{16a} \label{eqn:16a}\\ 
\textrm{s.t.} \quad & 0 \le p_{\text{se},n}^{(k)} \le P_{\text{se},n}^{\max}, \forall{k,n} \tag{16b} \label{eqn:16b}\\
& 0 \le p_{\text{cm},n}^{(k)}[t] \le P_{\text{cm},n}^{\max}, \forall{k,n,t} \tag{16c} \label{eqn:16c}\\
& 0 \le f_{n}^{(k)} \le f_{n}^{\max}, \forall{k,n} \tag{16d} \label{eqn:16d}\\
& E_{\text{total},n}^{\text{UAV}} \le E^{\max}, \forall{n} \tag{16e} \label{eqn:16e}\\
 (x_{n}[t+1] - x_{n}[t])^2 &+ (y_{n}[t+1] - y_{n}[t])^2 \leq (V_{\max} \delta_{t})^{2}, \forall{n,t}. \tag{16f} \label{eqn:16f}
\end{align}
\end{subequations}
\vspace{-20pt}
\\
\textit{\pmb{Sub-problem 2:}}
\vspace{-15pt}
\begin{subequations} 
\begin{align}
\min_{\begin{aligned}
    \text{\small $p_{\text{cm},\text{BS}}^{(k)}$, $f_{\text{BS}}^{(k)}$}
\end{aligned}} \quad & \sum_{k=1}^{K}\max_{n \in \mathcal{N}} \left\{T_{\text{se},n}^{(k)}+T_{\text{train},n}^{(k)}+T_{\text{cm},n}^{(k)}\right. \nonumber\\[-5pt] 
& \left.+L C_{\text{BS}}^{(k)} D_{\text{BS}}^{(k)}/f_{\text{BS}}^{(k)}+s_{g}/R_{\text{BS}}^{(k)}\right\} \tag{17a} \label{eqn:17a}\\ 
\textrm{s.t.} \quad & 0 \le p_{\text{cm},\text{BS}}^{(k)} \le P_{\text{cm},\text{BS}}^{\max}, \forall{k} \tag{17b} \label{eqn:17b}\\
& 0 \le f_{\text{BS}}^{(k)} \le f_{\text{BS}}^{\text{max}}, \forall{k}. \tag{17c} \label{eqn:17c}
\end{align}
\end{subequations}

It is clear that \textit{\pmb{sub-problem 1}} is  non-convex due to the objective function \eqref{eqn:16a} as well as constraint \eqref{eqn:16e}. Therefore, we now focus on convexification of \eqref{eqn:16a} and \eqref{eqn:16e}. 
\noindent

\subsection{Convexity of Sub-problem 1} \textit{\pmb{For the objective function,}} let us first introduce a slack variable $g$ such that:
\begin{align}
    \frac{s_{l}[t]}{B_{n} \log_{2}\left(1+\frac{\gamma_{0} p_{\text{cm},n}^{(k)}[t]}{\left(d_{n,\text{BS}}^{(k)}[t]\right)^2}\right)} \leq g. \label{eqn:22}
\end{align}
Now we introduce 3 more slack variables $z$, $\gamma$, and $\alpha$, and rewrite \eqref{eqn:22} as:
\begin{subnumcases}{\eqref{eqn:22}\Leftrightarrow}
   & $s_{l}[t] \leq g z$, \label{positive-subnum} \\
   & $B_{n} \log_{2}(1+\gamma) \geq z$, \label{zero-subnum} \\
   & $\frac{p_{\text{cm},n}^{(k)}[t]}{\alpha} \geq \gamma$, \label{negative-subnum} \\
   & $(x_{n}^{(k)}[t])^{2}+(y_{n}^{(k)}[t])^{2}+H^{2} \leq \alpha$. \label{negative-subnum2}
\end{subnumcases}

\noindent
It is evident that \eqref{eqn:22} is equivalent to the above set of equations. We now analyze the convexity of each inequality within it:
\\
\noindent
\uline{\textit{\pmb{\eqref{positive-subnum}:}}} We can equivalently write \eqref{positive-subnum} as $g z \geq s_{l}[t]$ which can be expressed as
\begin{align}
   &g z \geq s_{l}[t] \notag \\
   \Leftrightarrow \quad &\frac{1}{4} (g+z)^2 - \frac{1}{4} (g-z)^2 \geq s_{l}[t] \notag \\
   \Leftrightarrow \quad &\frac{1}{4} (g+z)^2 - s_{l}[t] \geq  \frac{1}{4} (g-z)^2. \label{eqn:25}
\end{align}
In \eqref{eqn:25}, the RHS is convex and thus we only need to approximate $(g+z)^2$. Using the first-order Taylor expansion, we have
\begin{align}
    (g+z)^2 \geq (g_{i}+z_{i})^2 + 2 (g_{i}+z_{i}) (g+z-g_{i}-z_{i}). \label{eqn:26}
\end{align}
By replacing \eqref{eqn:26} into \eqref{eqn:25}, we get
\begin{align}
    \frac{1}{4} &\left[(g_{i}+z_{i})^2 + 2 (g_{i}+z_{i}) (g+z-g_{i}-z_{i})\right] \notag \\
    &- s_{l}[t] \geq  \frac{1}{4} (g-z)^2. \label{eqn:27}
\end{align}
\vspace{-15pt}
\\
\uline{\textit{\pmb{\eqref{zero-subnum}:}}} By using this inequality \cite{4}
\begin{align}
    \ln(1+z) \ge \ln(1+z_{i}) + \frac{z_{i}}{z_{i}+1} - \frac{(z_{i})^{2}}{z_{i}+1} \frac{1}{z}, \label{eqn:28}
\end{align}
we approximate the LHS of \eqref{zero-subnum} as
\begin{align}
    \ln(1+\gamma_{i}) + \frac{\gamma_{i}}{\gamma_{i}+1} - \frac{(\gamma_{i})^{2}}{\gamma_{i}+1} \frac{1}{\gamma} \geq \frac{z \ln{2}}{B_{n}}. \label{eqn:29}
\end{align}
\\
\noindent
\uline{\textit{\pmb{\eqref{negative-subnum}:}}} We can equivalently write \eqref{negative-subnum} as
\begin{align}
    p_{\text{cm},n}^{(k)}[t] \geq \alpha \gamma. \label{eqn:30}
\end{align}
For $\alpha > 0$ and $\gamma > 0$, we apply SCA to approximate RHS of \eqref{eqn:30} as
\vspace{-10pt}
\begin{align}
    \alpha \gamma \leq \frac{1}{2} \frac{\gamma_{i}}{\alpha_{i}} \alpha^2 + \frac{1}{2} \frac{\alpha_{i}}{\gamma_{i}} \gamma^2,
\end{align}
where $\alpha_{i}$ and $\gamma_{i}$ are the feasible point of $\alpha$ and $\gamma$ at iteration $i$. Thus, \eqref{eqn:30} (equivalent of \eqref{negative-subnum}) can be convexified as
\vspace{-10pt}
\begin{align}
    p_{\text{cm},n}^{(k)}[t] \geq \frac{1}{2} \frac{\gamma_{i}}{\alpha_{i}} \alpha^2 + \frac{1}{2} \frac{\alpha_{i}}{\gamma_{i}} \gamma^2. \label{eqn:32}
\end{align}
\\
\noindent
\vspace{-28pt}

\uline{\textit{\pmb{\eqref{negative-subnum2}:}}} It is now clear that \eqref{negative-subnum2} is convex that can be directly solved by CVX.

\textit{\pmb{For the constraint \eqref{eqn:16e},}} we can equivalently write it as
\begin{align}
    &E_{\text{total},n}^{\text{UAV}} \le E^{\max}, \forall{k}, \notag \\
    \Leftrightarrow \, &\sum_{k=1}^{K}
    \left(E_{\text{se},n}^{(k)}+E_{\text{train},n}^{(k)}+E_{\text{cm},n}^{(k)}\right) \le E^{\max}, \forall{k}, \notag \\
    \Leftrightarrow \, &\sum_{k=1}^{K} \left(p_{\text{se},n}^{(k)} T_{u} D_{n}^{(k)} + S \zeta_{n}^{(k)} C_{n}^{(k)} D_{n}^{(k)} \left(f_{n}^{(k)}\right)^2\right. \notag + \\
    &\left. \sum_{t=1}^{T} s_{l}[t] p_{\text{cm},n}^{(k)}[t]/R_{n}^{(k)}[t] \right) \le E^{\max}, \forall{k}. \label{eqn:33}
\end{align}
From \eqref{eqn:33}, it is clear that the first and the second terms of the LHS are convex, while the third term is still non-convex. By replacing \eqref{eqn:22} into \eqref{eqn:33}, we get
\vspace{-10pt}
\begin{align}
    \sum_{k=1}^{K} &\Bigg(p_{\text{se},n}^{(k)} T_{u} D_{n}^{(k)} + S \zeta_{n}^{(k)} C_{n}^{(k)} D_{n}^{(k)} \left(f_{n}^{(k)}\right)^2 \notag \\
    &\quad  \sum_{t=1}^{T} g p_{\text{cm},n}^{(k)}[t]\Bigg) \le E^{\max}, \forall{k}. \label{eqn:34}
\end{align}

\vspace{-10pt}
Now, for $g > 0$ and $p_{\text{cm},n}^{(k)}[t] > 0$, we apply SCA to approximate $g p_{\text{cm},n}^{(k)}[t]$ as
\begin{align}
    g p_{\text{cm},n}^{(k)}[t] \leq \frac{1}{2} \frac{\mathop{p_{\text{cm},n}^{(k)}[t]}_{i}}{g_{i}} g_{i}^2 + \frac{1}{2} \frac{g_{i}}{\mathop{p_{\text{cm},n}^{(k)}[t]}_{i}} {p_{\text{cm},n}^{(k)}[t]}_{i}^2,
\end{align}
where $\mathop{p_{\text{cm},n}^{(k)}[t]}_{i}$ and $g_{i}$ are the feasible point of $p_{\text{cm},n}^{(k)}[t]$ and $g$ at iteration $i$. Thus, \eqref{eqn:33} (equivalent of constraint \eqref{eqn:16e}) can be convexified as
\begin{align}
    &\sum_{k=1}^{K} \Bigg(p_{\text{se},n}^{(k)} T_{u} D_{n}^{(k)} + S \zeta_{n}^{(k)} C_{n}^{(k)} D_{n}^{(k)} \left(f_{n}^{(k)}\right)^2 \notag \\
    &  \sum_{t=1}^{T} \left(\frac{1}{2} \frac{\mathop{p_{\text{cm},n}^{(k)}[t]}_{i}}{g_{i}} g_{i}^2 + \frac{1}{2} \frac{g_{i}}{\mathop{p_{\text{cm},n}^{(k)}[t]}_{i}} {p_{\text{cm},n}^{(k)}[t]}_{i}^2\right)\Bigg) \le E^{\max}, \forall{k}. \label{eqn:36}
\end{align}

\subsection{Convexity of Sub-problem 2} \textit{\pmb{For the objective function,}} similar to subproblem 1, let us introduce a slack variable such that:
\begin{align}
    \frac{s_{g}}{B_{\text{BS}} \log_{2}\left(1+\frac{\gamma_{0} p_{\text{cm},\text{BS}}^{(k)}}{\left(d_{\text{BS},n}^{(k)}\right)^2}\right)} \leq \Theta \label{eqn:37}
\end{align}

Sub-problem 2 can be equivalently re-written as
\begin{subequations} 
\begin{align}
\min_{\begin{aligned}
    \text{\small $p_{\text{cm},\text{BS}}^{(k)}$, $f_{\text{BS}}^{(k)}$}
\end{aligned}} \quad & \sum_{k=1}^{K}\max_{n \in \mathcal{N}} \left\{T_{\text{se},n}^{(k)}+T_{\text{train},n}^{(k)}+T_{\text{cm},n}^{(k)}\right. \nonumber\\[-5pt] 
& \left.+L C_{\text{BS}}^{(k)} D_{\text{BS}}^{(k)}/f_{\text{BS}}^{(k)}+\Theta\right\} \tag{33a} \label{eqn:33a}\\ 
\textrm{s.t.} \quad & 0 \le p_{\text{cm},\text{BS}}^{(k)} \le P_{\text{cm},\text{BS}}^{\max}, \forall{k} \tag{33b} \label{eqn:33b}\\
& 0 \le f_{\text{BS}}^{(k)} \le f_{\text{BS}}^{\text{max}}, \forall{k} \tag{33c} \label{eqn:33c}\\
& \frac{s_{g}}{B_{\text{BS}} \Theta} \leq \log_{2}\left(1+\frac{\gamma_{0} p_{\text{cm},\text{BS}}^{(k)}}{\left(d_{\text{BS},n}^{(k)}\right)^2}\right), \forall{k}. \tag{33d} \label{eqn:33d}
\end{align}
\end{subequations}
We see that \eqref{eqn:33a} is convex, while \eqref{eqn:33b}-\eqref{eqn:33c} are also convex. Now we focus on converting \eqref{eqn:33d} into convex one.
\\
\uline{\textit{\pmb{\eqref{eqn:33d}:}}} We use the same inequality in \eqref{eqn:28} to approximate RHS of \eqref{eqn:33d} as
\vspace{-5pt}
\begin{align}
    \frac{s_{g} \ln2}{B_{\text{BS}} \Theta} \leq \ln\left(1+\xi_{i}\right)+\frac{\xi_{i}}{\xi_{i}+1} - \frac{{\xi_{i}}^2}{\xi_{i}+1}.\frac{1}{\xi} \label{eqn:39}
\end{align}
where $\xi = \frac{\gamma_{0} \mathop p_{\text{cm},\text{BS}}^{(k)}}{\left(d_{\text{BS},n}^{(k)}\right)^2}$ and $\xi_{i} = \frac{\gamma_{0} \mathop{p_{\text{cm},\text{BS}}^{(k)}}_{i}}{\left(d_{\text{BS},n}^{(k)}\right)^2}$.
Based on above analysis, we are now ready to solve the following two equivalent convex sub-problems to obtain the solutions to the original problem (\pmb{Problem 1}) via standard optimization methods such as CVX.
\textit{\pmb{Sub-problem 1 (equivalent):}}
\begin{subequations} 
\begin{align}
\min_{\begin{aligned}
    & \text{\small $x_{n}^{(k)}[t]$, $y_{n}^{(k)}[t]$,} \\[-0.5ex]
    & \text{\small$p_{\text{se},n}^{(k)}$, $p_{\text{cm},n}^{(k)}[t]$,} \\[-0.5ex]
    & \text{\small $f_{n}^{(k)}$}
\end{aligned}} \quad & \sum_{k=1}^{K}\max_{n \in \mathcal{N}} \Bigg\{T_{u} D_{n}^{(k)}+\frac{S C_{n}^{(k)} D_{n}^{(k)}}{f_{n}^{(k)}} \nonumber \\[-30pt]
& +g+T_{\text{cp},\text{BS}}^{(k)}+T_{\text{dl},n}^{(k)}\Bigg\} \tag{35a} \label{eqn:35a}\\ 
\textrm{s.t.} \quad & \eqref{eqn:27},\eqref{eqn:29},\eqref{eqn:32},\eqref{negative-subnum2},\eqref{eqn:36},\eqref{eqn:16b}-\eqref{eqn:16d}, \eqref{eqn:16f}. \tag{35b} \label{eqn:35b}
\end{align}
\label{eqn:subproblem1}
\end{subequations}

\textit{\pmb{Sub-problem 2 (equivalent):}}
\begin{subequations} 
\begin{align}
\min_{\begin{aligned}
    \text{\small $p_{\text{cm},\text{BS}}^{(k)}$, $f_{\text{BS}}^{(k)}$}
\end{aligned}} \quad & \sum_{k=1}^{K}\max_{n \in \mathcal{N}} \Bigg\{T_{\text{se},n}^{(k)}+T_{\text{train},n}^{(k)}+T_{\text{cm},n}^{(k)} \nonumber\\[-5pt] 
& +\frac{L C_{\text{BS}}^{(k)} D_{\text{BS}}^{(k)}}{f_{\text{BS}}^{(k)}}+\Theta\Bigg\} \tag{36a} \label{eqn:36a}\\ 
\textrm{s.t.} \quad & \eqref{eqn:39},\eqref{eqn:33b}-\eqref{eqn:33c}. \tag{36b} \label{eqn:36b}
\end{align}
\label{eqn:subproblem2}
\end{subequations}
To summarize, we jointly solve the above two blocks to obtain the solutions for original \pmb{Problem 1}, as illustrated in Algorithm 1.
\makeatletter
\renewcommand{\Statex}{\item[]\hskip\ALG@thistlm}
\makeatother

\begin{algorithm}[t]
  \caption{SCA-based Joint Optimization Algorithm}
\begin{algorithmic}[1]
 \footnotesize
    \Statex \textbf{Input:} 
            \Statex Set the iteration index $i=0$;
            \Statex Initialize a feasible solution
            ${x_{n}^{(k)}[t]}_0$, ${y_{n}^{(k)}[t]}_0$, ${p_{\text{se},n}^{(k)}}_0$, ${p_{\text{cm},n}^{(k)}[t]}_0$, ${f_{n}^{(k)}}_0$, ${p_{\text{cm},\text{BS}}^{(k)}}_0$, ${f_{\text{BS}}^{(k)}}_0$ for Problem 1;
    \Statex \textbf{Repeat}
            \Statex Set $i \gets i+1$
            \Statex Solve convex \pmb{ Sub-problem 1} to update ${x_{n}^{(k)}[t]}_i$, ${y_{n}^{(k)}[t]}_i$, ${p_{\text{se},n}^{(k)}}_i$, ${p_{\text{cm},n}^{(k)}[t]}_i$, ${f_{n}^{(k)}}_i$;
            \Statex Solve convex \pmb{ Sub-problem 2} to update ${p_{\text{cm},\text{BS}}^{(k)}}_i$, ${f_{\text{BS}}^{(k)}}_i$;
    \Statex \textbf{Until} convergence. 
    \Statex \textbf{Output:}
            \Statex Optimal \pmb{${x_{n}^{(k)}[t]}^*$}, \pmb{${y_{n}^{(k)}[t]}^*$}, \pmb{${p_{\text{se},n}^{(k)}}^*$}, \pmb{${p_{\text{cm},n}^{(k)}[t]}^*$}, \pmb{${f_{n}^{(k)}}^*$}, \pmb{${p_{\text{cm},\text{BS}}^{(k)}}^*$}, \pmb{${f_{\text{BS}}^{(k)}}^*$}.
\end{algorithmic} 
\end{algorithm} 
\section{Simulation Results and Evaluation}
In this section, we conduct simulations  and evaluate the numerical performance of the proposed optimization scheme. All simulations were conducted in Matlab using YALMIP toolbox with the solver MOSEK. We have taken practical scenarios into account during the development of our simulation. The system bandwidth is set to 20 MHz \cite{6}. The maximum communication transmit power $P_{\text{cm},n}^{\text{max}}$ of UAV and $P_{\text{cm},\text{BS}}^{\text{max}}$ of BS are configured in the range of [5-25] dB and [15-35] dB, respectively. The maximum CPU cycle frequency of a UAV is configured as $f_{n}^{\text{mx}}$ = 2 GHz and that of the BS is configured as $f_{\text{BS}}^{\text{max}}$ = 10 GHz \cite{6}. The noise variance $\sigma^2$ is considered to be -80 dBm \cite{7}. The effective switched capacitance in local computation of UAV is $\zeta_{n}^{k} = 10^{-28}$ \cite{6}. The number of local iterations for each UAV is configured as $S$ = 15. To demonstrate the validity of our joint optimization method, we compare with two baselines: (i) Scheme 1-optimization for only UAVs and (ii) Scheme 2-optimization for only BS.
\begin{figure}[ht!]
    \centering
    \includegraphics[width=0.4\textwidth]{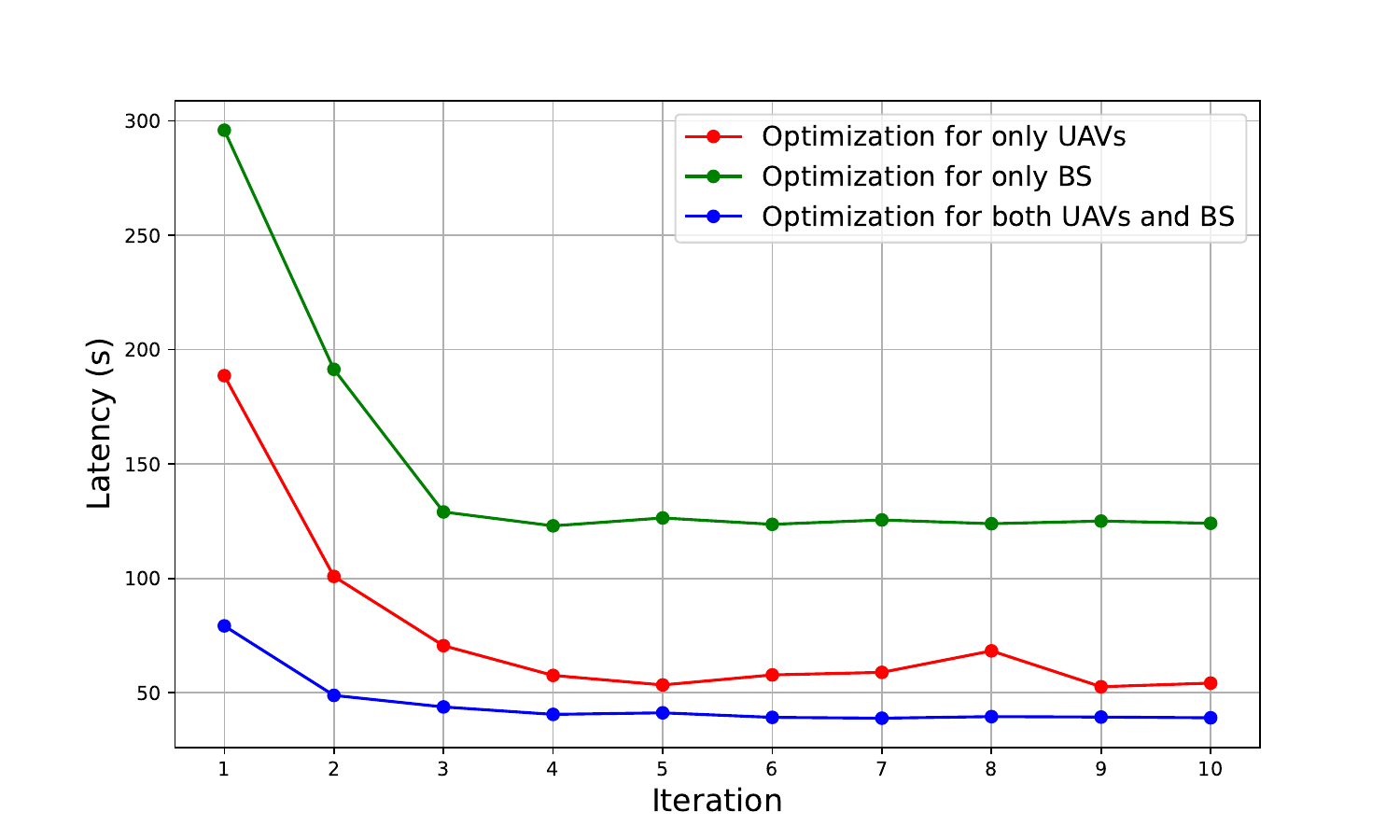} 
    \caption{\footnotesize Comparison of training performance among different schemes.}
    \label{fig1}
    \vspace{-7mm}
\end{figure}

Fig.~\ref{fig1} shows the convergence of our proposed convex optimization algorithm in terms of the system latency (second) versus the number of iterations. Comparing with Scheme 1 and Scheme 2, our approach achieves much lower system latency for the FL system, showing the merit of our joint optimization design.   The graph also reveals that our method maintains a steady latency after the fourth iteration, significantly outperforming the other two schemes in minimizing latency. Importantly, our proposed scheme results in 27.98\% less latency compared to Scheme 1 and 68.54\% less latency compared to Scheme 2.
\vspace{-15pt}
\begin{figure}[ht!]
    \centering
    \footnotesize
    \begin{subfigure}[t]{0.49\linewidth} 
        \centering
        \includegraphics[width=\linewidth]{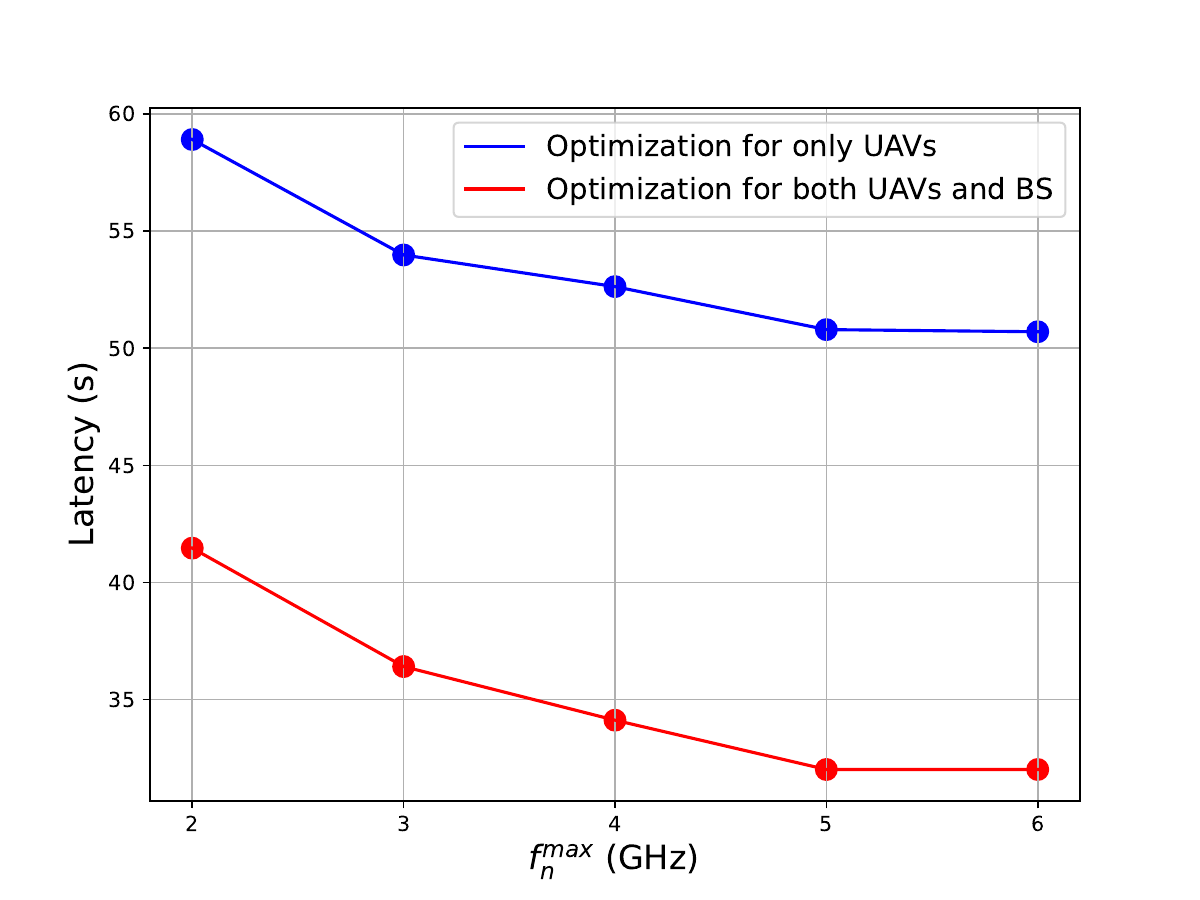}
        \caption{\footnotesize System latency versus maximum frequency of UAVs.}
        \label{fig2a}
    \end{subfigure}
    \hfill 
    \begin{subfigure}[t]{0.49\linewidth} 
        \centering
        \includegraphics[width=\linewidth]{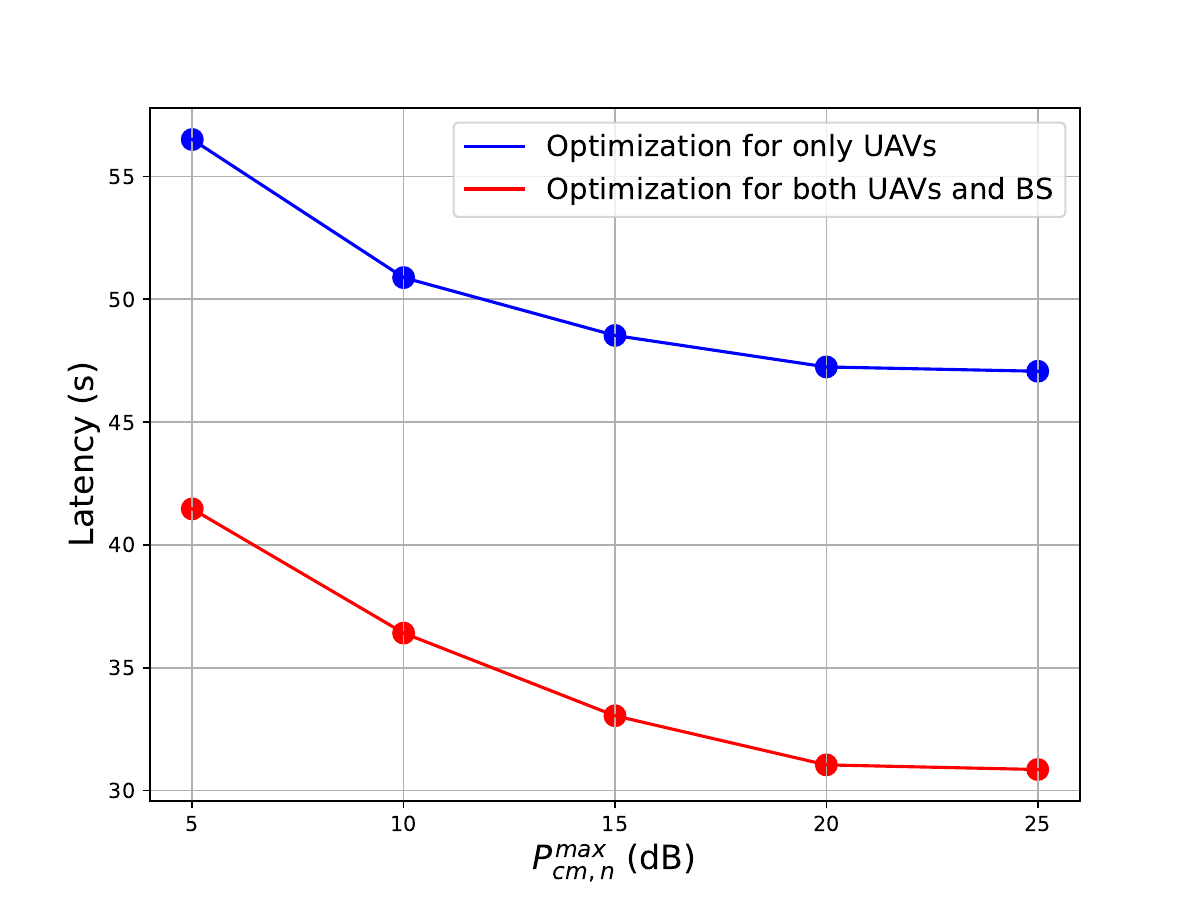}
        \caption{\footnotesize System latency versus maximum communication transmit power of UAVs.}
        \label{fig2b}
    \end{subfigure}
    \caption{\footnotesize Comparison of system latency with Scheme 1.}
    \label{fig:three_subs}
\end{figure}

\vspace{-10pt}
We also examine the latency performance of various schemes. Fig.~\ref{fig2a} shows the latency (second) against the maximum CPU computation capability (GHz) of a UAV, comparing the performance of our proposed algorithm with that of Scheme 1. Higher frequencies typically result in higher data rates, which decrease latency by enabling quicker transmission and reception of packets. Both schemes demonstrate reduction in latency as the maximum frequency of UAVs rises, but the proposed scheme shows a significant 36.85\% decrease in latency compared to Scheme 1. The enhanced performance of the proposed algorithm is attributed to its capability to dynamically adapt to network conditions by optimizing parameters for both UAVs and BS, leading to more efficient resource utilization and lower latency.

Fig.~\ref{fig2b} illustrates the latency (second) versus the maximum communication transmit power of a UAV, comparing our proposed scheme with Scheme 1. Higher transmit power reduces propagation delays, resulting in quicker signal transmission and increased data rates. Our proposed scheme achieves 34.44\% lower latency than Scheme 1, even though both schemes show decreased latency with higher maximum transmit power. By optimizing resource allocation for both UAVs and the BS, our joint optimization method can achieve better latency savings for robust FL training.
\begin{figure}[ht!]
    \centering
    \footnotesize
    \begin{subfigure}[t]{0.49\linewidth} 
        \centering
        \includegraphics[width=\linewidth]{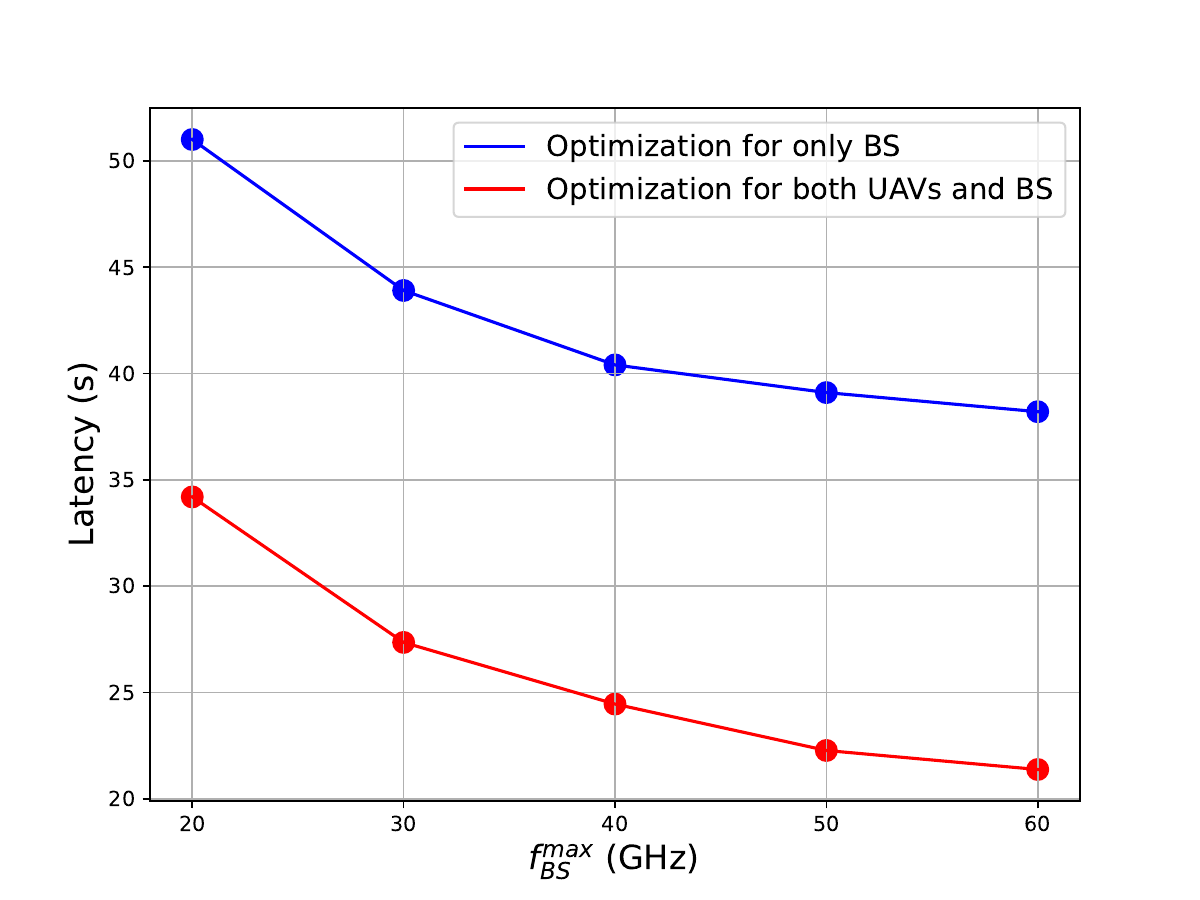}
        \caption{\footnotesize System latency versus maximum frequency of BS.}
        \label{fig3a}
    \end{subfigure}
    \hfill 
    \begin{subfigure}[t]{0.49\linewidth} 
        \centering
        \includegraphics[width=\linewidth]{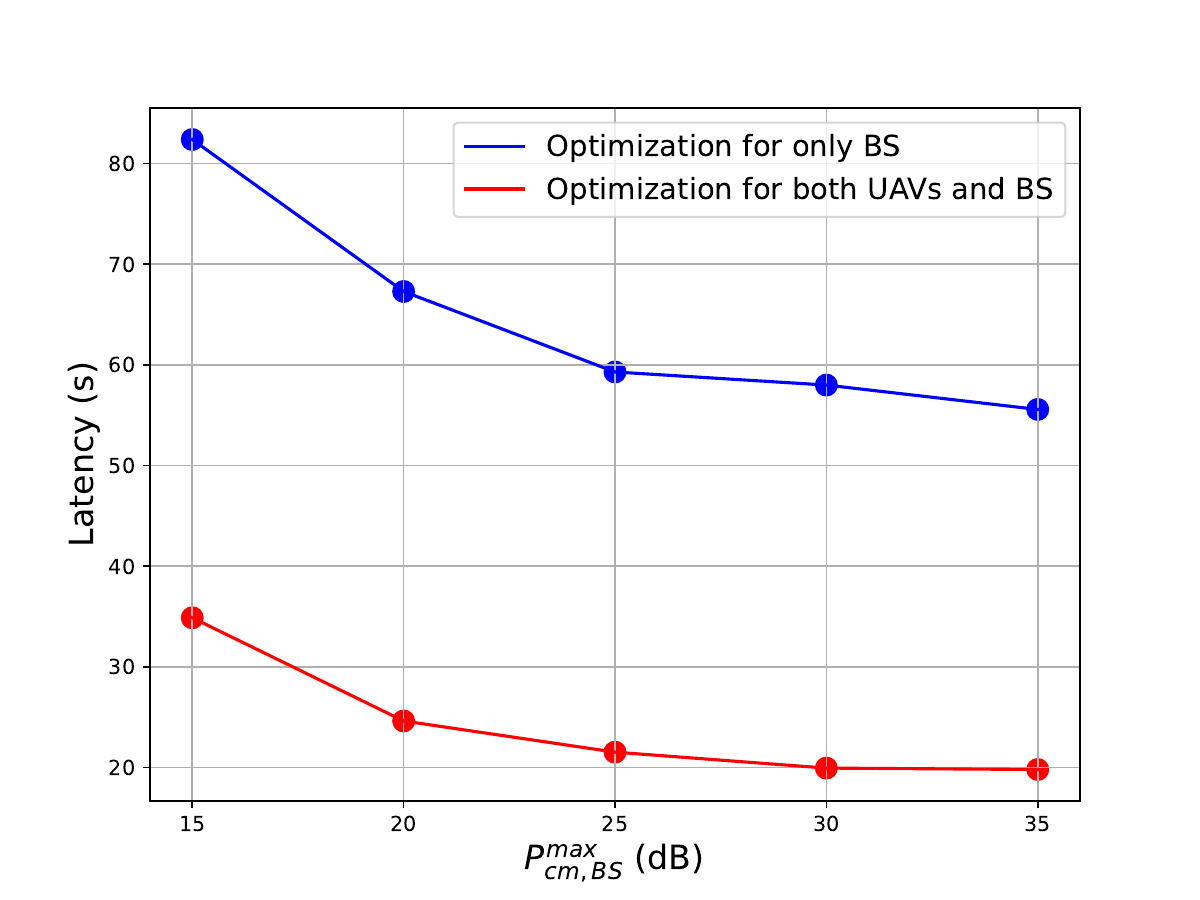}
        \caption{\footnotesize System latency versus maximum BS's transmit power.}
        \label{fig3b}
    \end{subfigure}
    \caption{\footnotesize Comparison of system latency with Scheme 2.}
    \label{fig5}
    \vspace{-7mm}
\end{figure}

Fig.~\ref{fig3a} compares our proposed algorithm with Scheme 2 demonstrating latency  versus maximum CPU computation capability (GHz) of BS. Both schemes see a gradual decrease in latency as BS frequency increases, with our proposed scheme outperforming Scheme 2 achieving 44.04\% lower latency. Finally, Fig.~\ref{fig3b} presents latency  versus maximum communication transmit power of BS, again highlighting the comparison between our proposed method and Scheme 2. Both schemes experience reduced latency with higher transmit power, yet our proposed method surpasses Scheme 2 by achieving 64.37\% lower latency.
\section{Conclusion}
This paper has studied a latency optimization problem for UAV-enabled FL system with ISAC, by jointly optimizing resource allocation and trajectory of UAVs and resource allocation of the BS. The formulated problem is non-convex and thus challenging to solve. Therefore, we have proposed an efficient iterative optimization approach via the BCD and SCA techniques to obtain optimal solutions. Simulation results have demonstrated that our joint optimization approach effectively decreases the system latency by up to 68.54\% compared to baseline methods. 
\vspace{-3pt}
\bibliographystyle{IEEEtran}
\bibliography{Main-Bibliography}

\end{document}